\documentclass[runningheads]{llncs}

\usepackage[T1]{fontenc}

\usepackage{notoccite}

\usepackage{graphicx}
\usepackage{comment}
\usepackage{svg}
\usepackage{cite}
\usepackage{color}
\usepackage{hyperref}
\usepackage{float}
\usepackage{multirow}
\usepackage{multirow}

\begin{document}

\title{Brain Tumor Segmentation in Sub-Sahara Africa with Advanced Transformer and ConvNet Methods: Fine-Tuning, Data Mixing and Ensembling}
\titlerunning{Brain Tumor Segmentation in Sub-Sahara Africa}

\author{Toufiq Musah\inst{1, 19} \and
Chantelle Amoako-Atta\inst{2} \and
John Amankwaah Otu\inst{3} \and
Lukman E. Ismaila\inst{6, 20} \and
Swallah Alhaji Suraka\inst{4} \and
Oladimeji Williams\inst{5} \and
Isaac Tigbee\inst{8} \and
Kato Hussein Wabbi\inst{7} \and
Samantha Katsande\inst{9} \and
Kanyiri Ahmed Yakubu\inst{8} \and
Adedayo Kehinde Lawal\inst{10} \and
Anita Nsiah Donkor\inst{8} \and
Naeem Mwinlanaah Adamu\inst{8} \and
Adebowale Akande\inst{11} \and
John Othieno\inst{12} \and
Prince Ebenezer Adjei\inst{1, 19} \and
Zhang Dong\inst{14} \and
Confidence Raymond\inst{16} \and
Udunna C. Anazodo\inst{15,16} \and
Abdul Nashirudeen Mumuni\inst{8} \and 
Adaobi Chiazor Emegoakor\inst{17} \and
Chidera Opara\inst{15} \and
Maruf Adewole\inst{15} \and
Richard Asiamah\inst{18}
}

\authorrunning{T. Musah et al.}

\institute{
Department of Computer Engineering, Kwame Nkrumah University of Science and Technology, Kumasi, Ghana.\\ \and
Department of Mathematics, University of Ghana, Accra, Ghana.\\ \and
Department of Physics, Kwame Nkrumah University of Science and Technology, Kumasi, Ghana.\\ \and
Department of Medical Imaging, University of Health and Allied Sciences, Ho, Ghana.\\ \and
Department of Electrical and Electronics Engineering, University of Lagos, Akoka, Nigeria.\\ \and
F.M. Kirby Research Center for Functional Brain Imaging at Kennedy Krieger Institute, MD USA.\\ \and
Makerere University Biomedical Research Centre, School of Biomedical Sciences, College of Health Sciences, Makerere University, Uganda.\\ \and
Department of Medical Imaging, University for Development Studies, Tamale, Ghana.\\ \and
Action on Preeclampsia Ghana (APEC-GH), Accra, Ghana.\\ \and
Department of Radiology, Lagos State University Teaching Hospital, Ikeja, Nigeria.\\ \and
Department of Electronic and Electrical Engineering, Obafemi Awolowo University, Ile-Ife, Osun State, Nigeria.\\ \and
Department of ICT and Biostatistics, Ernest Cook UltraSound Research and Education Institute, Mengo, Kampala, Uganda.\\ \and
Department of Radiology, Komfo Anokye Teaching Hospital, Kumasi, Ghana.\\ \and
Department of Electrical \& Computer Engineering, University of British Columbia, Vancouver, BC, Canada.\\ \and
Medical Artificial Intelligence Lab, Lagos, Nigeria.\\ \and
Montreal Neurological Institute, McGill University, Montreal, Canada.\\ \and
Nnamdi Azikiwe University Teaching Hospital, Nnewi, Nigeria \\ \and
Department of Biomedical Engineering, University of Ghana, Accra.\\ \and
Global Health and Infectious Disease Group, Kumasi Centre for Collaborative Research in Tropical Medicine, Kumasi, Ghana.\\ \and
Johns Hopkins University, Baltimore, MD, USA.\\
\email{toufiqmusah32@gmail.com}
}


\maketitle              

\begin{abstract}
Brain tumors are among the deadliest cancers worldwide, with particularly devastating impact in Sub-Saharan Africa (SSA) where limited access to medical imaging infrastructure and expertise often delays diagnosis and treatment planning. Accurate brain tumor segmentation is crucial for treatment planning, surgical guidance, and monitoring disease progression, yet manual segmentation is time-consuming and subject to inter-observer variability. Recent advances in deep learning, based on Convolutional Neural Networks (CNNs) and Transformers have demonstrated significant potential in automating this critical task. This study evaluates three state-of-the-art architectures, SwinUNETR-v2, nnUNet, and MedNeXt for automated brain tumor segmentation in multi-parametric Magnetic Resonance Imaging (MRI) scans. We trained our models on the BraTS-Africa 2024 and BraTS2021 datasets, and performed validation on the BraTS-Africa 2024 validation set. We observed that training on a mixed dataset (BraTS-Africa 2024 and BraTS2021) did not yield improved performance on the SSA validation set in all tumor regions compared to training solely on SSA data with well-validated methods. Ensembling predictions from different models also lead to notable performance increases. Our best-performing model, a finetuned MedNeXt, achieved an average lesion-wise Dice score of 0.84, with individual scores of 0.81 (enhancing tumor), 0.81 (tumor core), and 0.91 (whole tumor). While further improvements are expected with extended training and larger datasets, these results demonstrate the feasibility of deploying deep learning for reliable tumor segmentation in resource-limited settings. We further highlight the need to improve local data acquisition protocols to support the development of clinically relevant, region-specific AI tools.

\end{abstract}

\section{Introduction}
Glioblastomas, commonly referred to as brain tumors, are among the most aggressive and proliferative forms of cancer with a five-year survival rate of only 6.9\% as reported by the National Brain Tumor Society \cite{NBTS_Glioblastoma}. In Sub-Saharan Africa (SSA), where cases were once considered rare, recent findings have challenged this notion, suggesting that the perceived rarity is largely due to under-reporting \cite{kanmounye2022adult, odeku1972tumors}. This under-reporting has been linked to the lack of tumor diagnostic tools centralized tumor boards \cite{mbi2021need, establish}, a consequence of resource limitations in brain tumor research across the continent \cite{ngulde2015improving, ismaila2024afribiobank}. In 2022 alone, about 10,000 new glioblastoma cases were recorded in the SSA region, with a mortality rate of 84\%, according to the World Health Organization’s International Agency for Research on Cancer (IARC) \cite{globocan2019}. This high mortality rate reflects the gaps in early detection, access to diagnostic tooling, and effective treatment, highlighting the need for context-specific research and improved clinical strategies in the region \cite{establish, anazodo2023framework}.

Accurately delineating brain tumor regions on MRI provides vital information on tumor size, location, and extent, directly informing surgical approach and radiation treatment plans \cite{kouli2022automated}. In current practice, tumor regions in neuroimages are outlined manually by expert radiologists. While considered the reference standard, this process is time-consuming, and may suffer from inter- and intra-observer variability based on the annotating radiologist's level of expertise, and other compounding factors \cite{bo2017intra, jungo2018effect}. Automated methods, including deep learning, have the potential to address the stated challenges in the segmentation of brain tumors \cite{dorfner2025review}. However, resource constraints in SSA and other LMICs introduce additional hurdles, including suboptimal image quality and resolution \cite{adewole2023brain} due to a lack of onsite imaging expertise and low-quality imaging equipment \cite{anazodo2023framework}. A contributing constraint is the lack of sufficient annotated datasets for training robust deep learning algorithms designed to support diagnosis in SSA. For instance, in the Brain Tumor Segmentation (BraTS) challenge, the primary task of automated segmentation of glioma (BraTS-GLI) includes over 1,250 training cases \cite{baid2021rsna}, whereas the Africa-specific task (BraTS-Africa) has only 60 training cases \cite{adewole2023brain}.

Deep learning-based methods have demonstrated superior performances in biomedical image segmentation, with the U-Net architecture by \cite{ronneberger2015u} being the most widely adopted. Several variants of the U-Net framework emerged, primarily ConvNet based, including the attention U-Net \cite{oktay2018attention}, which make use of attention gates to enable the model to implicitly suppress irrelevant regions. 3D U-Net \cite{cciccek20163d} and  V-Net \cite{milletari2016v} introduced 3D volumetric convolutions for 3D segmentation, and nnUNet \cite{isensee2021nnu} has also been developed with the ability to adapt to a given segmentation dataset, and has so far demonstrated the best performances out-of-the-box on several tasks \cite{isensee2024nnu}. Recently, transformer based methods have gained traction, especially Swin UNETR \cite{hatamizadeh2021swin} which composed of an encoder that partitions the input image into local windows and applies a shifting window self-attention mechanism to capture both local and global contextual information \cite{liu2021swin}, and a ConvNet decoder. An updated Swin UNETR-v2 built on this, introducing stagewise convolutions within the transformer encoder to improve upon the model's inductive bias. Lastly, MedNeXt \cite{roy2023MedNeXt} built on the ConvNeXt architecture \cite{liu2022convnet}, a fully convolutional paradigm, adopting design principles from the shifting window transformer \cite{liu2021swin}.

In this work, we applied state-of-the-art transformer- and ConvNet- based methods to automatically segment brain tumors on MRI scans of SSA patients. Taking into account region-specific challenges, such as suboptimal imaging quality and limited availability of annotated data, we evaluate these methods to identify the most suitable approaches for addressing these constraints.  We further investigate the impact of training with non-homogeneous datasets from diverse sources, including BraTS 2021 \cite{baid2021rsna}, and assess the effectiveness of fine-tuning models pretrained on these external datasets. We employed the STAPLE (Simultaneous Truth and Performance Level Estimation) ensembling technique on selected predictions, demonstrating that the best-performing outputs from multiple models can be effectively combined to achieve improved segmentation results on limited training datasets.

\section{Materials and Methods}

\subsection{Data}
The primary dataset utilized in this study is the BraTS-Africa 2024 dataset, which includes multi-parametric MRI scans from 95 individuals consisting of diverse cases relevant to the target population in SSA \cite{adewole2023brain}. 60 samples are designated for training and 35 samples for validation. Data preprocessing by the challenge organizers prior to releasing the dataset included patient data anonymization, skull stripping, and co-registering the MRI scans to a common anatomical template. 


\subsection{Model Selection}
\subsubsection{SwinUNETR-v2} SwinUNETR-v2 (Shifting Window U-Net Transformer) \cite{he2023swinunetr} is built upon the original SwinUNETR \cite{hatamizadeh2021swin}, introducing architectural changes that led to slight benchmark improvements in various biomedical imaging segmentation tasks. The SwinUNETR architecture consists of a transformer-based encoder that performs downsampling and feature extraction at multiple stages, connected to a convolutional decoder that concatenates and upsamples features to produce the final segmentation output. A purely transformer-based encoder may offer the advantage of global context awareness through the attention mechanism, but lacks the inductive bias of convolutions, which has been shown to be beneficial for medical image segmentation tasks, as demonstrated in models like the U-Net \cite{ronneberger2015u}. To harness the strengths of both approaches, SwinUNETR-v2 introduces stagewise convolutional modules after each transformer downsampling stage in the encoder, thereby reintroducing inductive bias into the feature extraction process.

\subsubsection{nnUNet} The nnUNet (no new U-Net) framework provides an out-of-the-box tool for biomedical image segmentation \cite{isensee2021nnu}, which we have previously demonstrated to perform admirably in neuroimaging segmentation tasks \cite{musah2024automated}. In this study, we specifically make use of the new Residual Encoder presets \cite{isensee2024nnu}. It introduced a stronger residual encoder for better feature extraction in the earlier stages of the U-Net.

\subsubsection{MedNeXt} MedNeXt is a transformer-inspired fully-ConvNeXt encoder-decoder U-Net, specially made for 3D biomedical image segmentation \cite{roy2023MedNeXt}. Building on the ConvNeXt architecture introduced by Liu et al. \cite{liu2022convnet}, it employs architectural elements originally devised for Shifting Window Transformers, including larger convolutional kernel sizes and an inverted bottleneck design, improving the network's ability to capture global representation. We specifically selected the medium variant of MedNeXt with a kernel size of $3 \times 3 \times 3$ and trained it using the nnUNet training pipeline \cite{isensee2021nnu}.

\subsection{STAPLE Ensembling}
Simultaneous Truth and Performance Level Estimation (STAPLE) was explored in a limited capacity as an ensembling strategy, where predictions from both nnUNet and MedNeXt were combined. STAPLE works by estimating a probabilistic ground truth segmentation from a collection of candidate segmentations, weighting each one based on its estimated performance \cite{warfield2004simultaneous}. In our case, STAPLE was applied to segmentations from the validation split, using outputs from both models to generate a consensus segmentation that showed improved performance over individual predictions. The algorithm estimates the true label $T_v$ at each voxel $v$, by maximizing the posterior probability:
\[
P(T_v = 1 \mid D_v)
\]

where $D_v$ represents the set of predicted segmentationa from MedNeXt and nnUNet at voxel, producing a voxel-wise probability map estimating the likelihood of each voxel being part of a tumor, based on inter-model agreement.

\subsection{Experiments}
We conducted several experiments utilizing the BraTS-Africa dataset, alongside the publicly available BraTS2021 dataset, to explore different training strategies and evaluate their impact on segmentation performance. Initially, each model was trained from scratch exclusively on the BraTS-Africa dataset.

We also explored a \emph{data mixing} approach, combining the original 60 BraTS-Africa samples with 200 randomly selected samples from the BraTS2021 dataset. This experiment aimed to assess whether incorporating additional BraTS2021 data could improve segmentation performance. Contrary to expectations, initial evaluations suggested a slight performance drop, particularly with nnUNet, likely due to differences in data quality and inherent noise levels between the two datasets. Such quality issues in the BraTS-Africa dataset have been previously noted \cite{adewole2023brain, anazodo2023framework}. This is further discussed in our results section.

Additionally, for the SwinUNETR model, we evaluated the effect of \emph{pretraining}, where the model was initially trained on the BraTS2021 dataset and subsequently fine-tuned on BraTS-Africa data. nnUNet and MedNeXt have exhibited strong out-of-the-box performance without extensive pretraining. We explored fine-tuning these models further on the BraTS-Africa dataset, initializing training with weights obtained from initial training on the same dataset. This fine-tuning step yielded additional performance improvements.

\section{Results}

\begin{figure}[ht]
    \centering
    \includegraphics[width=1.0\textwidth]{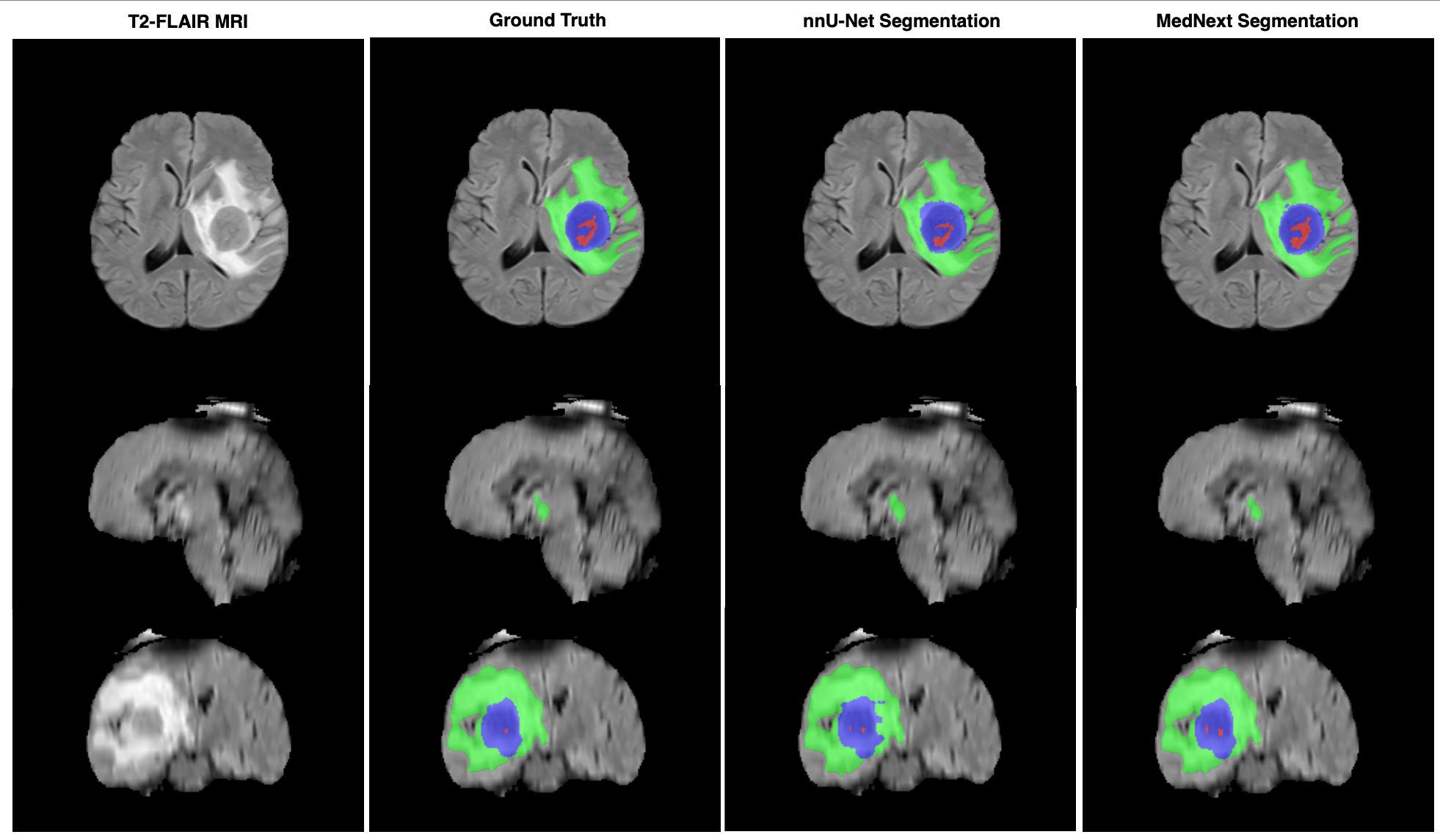}
    \caption{An example of the predicted tumor sub-regions from the nnUNet and MedNeXt models, shown in case 055. (Tumor Core-Red, Enhancing Tumor-Blue, and Whole Tumor-Green.}
    \label{fig:results}
\end{figure}


\subsection{Evaluation Metrics}
Our solution is evaluated using the Dice Similarity Coefficient (DSC), Hausdorff Distance at the 95th percentile (HD95), as well as lesion-wise Dice and lesion-wise HD95 \cite{menze2015multimodal}. These metrics are computed across the following tumor subregions, as defined in the 2024 BraTS challenge: tumor core (TC), enhancing tumor (ET), and whole tumor (WT).

\subsection{Segmentation Performance}
Tables~\ref{tab:all-dice-scores} and~\ref{tab:all-hd95-scores} summarize the segmentation performance of the nnUNet and MedNeXt models across different training setups. Results are reported for both lesion-wise and legacy variants of the Dice score and HD95 metric, measured on the tumor sub-regions, along with the corresponding averages.

For nnUNet, three configurations were evaluated: training on BraTS-Africa (nn$_{SSA}$), fine-tuning on BraTS-Africa (nn$_{SSA-FT}$), and data mixing with BraTS2021 (nn$_{Mix}$). MedNeXt was evaluated using the same BraTS-Africa data (M$_{SSA}$) and its fine-tuned variant (M$_{SSA-FT}$).

\begin{table}[ht]
\centering
\setlength{\tabcolsep}{4pt}
\renewcommand{\arraystretch}{1.15}
 
\caption{Dice scores for nnUNet (nn) and MedNeXt (M) models. Lesion-wise and legacy variants are reported per region and as an average. Best shown in \textbf{bold}.}
\label{tab:all-dice-scores}
\begin{tabular}{l|ccc|c|ccc|c}
\hline
\multirow{2}{*}{\textbf{Model}} & \multicolumn{4}{c|}{\textbf{Lesion-wise Dice}} & \multicolumn{4}{c}{\textbf{Legacy Dice}} \\
\cline{2-9}
 & ET & TC & WT & Avg & ET & TC & WT & Avg \\
\hline
nn$_{SSA}$     & 0.797 & 0.786 & 0.846 & 0.810 & 0.850 & 0.853 & 0.913 & 0.872 \\
nn$_{SSA-FT}$  & \textbf{0.805} & \textbf{0.794} & 0.853 & \textbf{0.817} & \textbf{0.871} & \textbf{0.873} & \textbf{0.923} & \textbf{0.889} \\
nn$_{Mix}$     & 0.759 & 0.766 & \textbf{0.872} & 0.799 & 0.836 & 0.844 & 0.904 & 0.861 \\
\hline
M$_{SSA}$      & 0.793 & 0.786 & 0.893 & 0.824 & \textbf{0.878} & \textbf{0.876} & 0.924 & \textbf{0.893} \\
M$_{SSA-FT}$   & \textbf{0.811} & \textbf{0.806} & \textbf{0.907} & \textbf{0.841} & 0.869 & 0.869 & \textbf{0.924} & 0.887 \\
\hline
\end{tabular}
\end{table}
\begin{table}[ht]
\centering
\setlength{\tabcolsep}{4pt}
\renewcommand{\arraystretch}{1.15}
 
\caption{HD95 scores (mm) for nnUNet (nn) and MedNeXt (M) models. Lesion-wise and legacy variants are reported per region and as an average. Lower is better. Best shown in \textbf{bold}.}
\label{tab:all-hd95-scores}
\begin{tabular}{l|ccc|c|ccc|c}
\hline
\multirow{2}{*}{\textbf{Model}} & \multicolumn{4}{c|}{\textbf{Lesion-wise HD95 ($\downarrow$)}} & \multicolumn{4}{c}{\textbf{Legacy HD95 ($\downarrow$)}} \\
\cline{2-9}
 & ET & TC & WT & Avg & ET & TC & WT & Avg \\
\hline
nn$_{SSA}$     & 42.717 & 49.288 & 39.053 & 43.686 & 24.944 & 26.819 & 6.743  & 19.502 \\
nn$_{SSA-FT}$  & \textbf{37.213} & \textbf{43.913} & 35.188 & \textbf{38.771} & \textbf{14.262} & \textbf{16.187} & 7.422 & \textbf{12.623} \\
nn$_{Mix}$     & 53.574 & 54.706 & \textbf{14.953} & 41.078 & 26.088 & 27.091 & \textbf{5.768} & 19.649 \\
\hline
M$_{SSA}$      & \textbf{45.470} & \textbf{48.889} & \textbf{14.157} & \textbf{36.172} & \textbf{13.615} & \textbf{15.797} & \textbf{5.329}  & \textbf{11.580} \\
M$_{SSA-FT}$   & 46.344 & 49.539 & 19.325 & 38.402 & 24.012 & 26.068 & 14.485 & 21.521 \\
\hline
\end{tabular}
\end{table}

\noindent SwinUNETR-v2 was evaluated only through the a held out subset of 12 cases from the training dataset. A complete Synapse submission could not be submitted within the project timeline. The resulting dice scores are summarized in Table~\ref{tab:swinunetr-v2}.  

\begin{table}[ht]
\centering
\setlength{\tabcolsep}{6pt}
\renewcommand{\arraystretch}{1.15}
 
\caption{In-training validation Dice scores for SwinUNETR-v2 on 12 Cases.}
\label{tab:swinunetr-v2}
\begin{tabular}{l|ccc|c}
\hline
\textbf{Dataset} & \textbf{ET} & \textbf{TC} & \textbf{WT} & \textbf{Avg} \\
\hline
BraTS-Africa               & 0.623 & 0.680 & 0.752 & 0.720 \\
BraTS-Mix               & 0.796 & 0.848 & 0.874 & 0.839 \\
BraTS2021               & \textbf{0.847} & \textbf{0.917} & 0.888 & \textbf{0.884} \\
BraTS-Africa-FT on BraTS2021  & 0.813 & 0.881 & \textbf{0.923} & 0.872 \\
\hline
\end{tabular}
\end{table}

\subsubsection{STAPLE Ensemble} 

We further explored a STAPLE ensemble combining predictions from nn$_{SSA}$, nn$_{SSA-FT}$, and M$_{SSA}$. The resulting lesion-wise dice scores are 0.805, 0.791, 0.895, for ET, TC and WT respectively. The ensemble achieved improved overall performance, with an average Dice score of 0.830 across the three tumor sub-regions, outperforming each of the individual models in isolation.


\section{Discussion}
Fine-tuning with SSA-specific data consistently improved performance across all tested models. Both MedNeXt and nnUNet showed notable gains after fine-tuning, with the fine-tuned models (M${SSA-FT}$ and nn${SSA-FT}$) achieving the highest average Dice scores and the lowest HD95 distances, respectively. Interestingly, performance on whole tumor (WT) segmentation remained strong across several configurations, including nn$_{Mix}$, suggesting that WT may be less sensitive to domain-specific variation than subregions like the enhancing tumor (ET) or tumor core (TC).

Contrary to expectations, mixing SSA data with samples from the BraTS2021 dataset did not lead to improved performance. The nn$_{Mix}$ model, while competitive on WT segmentation, underperformed on ET and TC compared to its fine-tuned SSA-only counterpart. This result aligns with prior observations about domain shift and the potential negative effects of mixing heterogeneous datasets. Specifically, the presence of varying image quality, annotation styles, and scanner characteristics in SSA data may introduce inconsistencies that disrupt learned feature representations when directly combined with higher-quality, curated datasets like BraTS2021. Notably, nnUNet’s internal regularization mechanisms helped mitigate performance degradation in mixed settings, but still did not surpass the SSA-fine-tuned baseline. These findings suggest that datasets like BraTS2021 may be better leveraged for pretraining, allowing models to capture generalizable features before adapting to the SSA domain via targeted fine-tuning.

MedNeXt also demonstrated strong baseline performance, with the out-of-the-box model (M${SSA}$) achieving competitive legacy Dice and HD95 scores, particularly for WT. However, fine-tuning (M${SSA-FT}$) further enhanced lesion-wise performance, reaffirming the value of domain adaptation even for architectures designed with cross-domain robustness in mind.

Ensembling via the STAPLE method, incorporating predictions from nn${SSA}$, nn${SSA-FT}$, and M$_{SSA}$, yielded an average lesion-wise Dice of 0.830, surpassing the performance of each model individually. This highlights the advantage of ensemble strategies in data-constrained environments, where combining model outputs can compensate for individual weaknesses and enhance overall robustness.

Due to resource limitations, SwinUNETR-v2 was only evaluated using in-training validation. Although the fine-tuned SwinUNETR demonstrated competitive Dice performance, the absence of external validation precludes direct comparison with other models. Moreover, its poor generalization on the BraTS-Africa dataset alone is consistent with known limitations of transformer-based architectures, which typically require large-scale datasets for effective training and generalization \cite{vit-understand}.

Collectively, these findings emphasize the importance of domain-specific fine-tuning, the risks of direct dataset mixing across heterogeneous sources, and the utility of lightweight ensembling methods in low-resource environments. This work contributes actionable insights for enhancing the robustness, generalizability, and practical deployment of segmentation models in SSA and similarly underrepresented medical imaging settings.

\section{Conclusion}
In summary, this study emphasizes the importance of developing segmentation solutions using data collected from the environments in which they are intended to be applied. Models trained or fine-tuned on Sub-Saharan African MRI data consistently outperformed those relying solely on mixed or external datasets, particularly in the presence of domain shifts. While established architectures like nnUNet and MedNeXt demonstrated strong baseline performance, further improvements were observed through fine-tuning and lightweight ensembling. These findings support the view that robust and context-relevant solutions requires direct engagement with local data, rather than relying exclusively on datasets curated in unrelated settings.

\section*{Acknowledgement}
The authors would like to thank all faculty and instructors of the Sprint AI Training for African Medical Imaging Knowledge Translation (SPARK) Academy\footnote{\url{https://www.cameramriafrica.org/spark}} 2024 Summer School on Deep Learning in Medical Imaging for providing valuable insights into brain tumor pathology, which informed the research presented here. Special thanks to Linshan Liu for administrative support throughout the SPARK Academy training and capacity-building activities from which the authors greatly benefited. The authors also acknowledge the contributions of Akwasi Asare, Edward Amenyaglo, and Edifon Jimmy. Computational infrastructure support was provided by the Digital Research Alliance of Canada (The Alliance), with additional knowledge translation support from the McGill University Doctoral Internship Program through the student exchange component of the SPARK Academy. The authors are further grateful to McMedHacks for delivering foundational training in Python programming for medical image analysis as part of the 2024 SPARK Academy curriculum. This research was supported by the Lacuna Fund for Health and Equity (PI: Udunna Anazodo, grant number 0508-S-001) and the Natural Sciences and Engineering Research Council of Canada (NSERC) Discovery Launch Supplement (PI: Udunna Anazodo, grant number DGECR-2022-00136).

\bibliographystyle{unsrt}
\bibliography{references}

\end{document}